\documentclass[11pt,letterpaper]{JHEP3}
\usepackage{graphics}
\usepackage{epsfig}
\usepackage{amssymb}
\usepackage{stmaryrd}
\usepackage{amsmath}
\usepackage{amsfonts}
\usepackage{mathrsfs}
\usepackage{graphicx}
\def\be{\begin{equation}}
\def\ee{\end{equation}}
\def\ba{\begin{eqnarray}}
\def\ea{\end{eqnarray}}

\title{Note on Zero Temperature Holographic Superfluids}
\author{Minyong Guo\\
Department of Physics, Beijing Normal University, Beijing, 100875, China
\email{minyongguo@mail.bnu.edu.cn}}

\author{Shanquan Lan\\
Department of Physics, Beijing Normal University, Beijing 100875, China\\
\email{shanquanlan@mail.bnu.edu.cn}}

\author{Chao Niu\\
School of Physics and Chemistry, Gwangju Institute of Science and Technology, Gwangju 500-712, Korea\\
\email{chaoniu09@gmail.com}}

\author{Yu Tian\\
School of Physics, University of Chinese Academy of Sciences,
Beijing 100049, China\\
State Key Laboratory of Theoretical Physics, Institute of
Theoretical Physics, Chinese Academy of Sciences, Beijing 100190,
China\\
\email{ytian@ucas.ac.cn}}

\author{Hongbao Zhang\\
Department of Physics, Beijing Normal University, Beijing, 100875, China\\
Theoretische Natuurkunde, Vrije Universiteit Brussel, and The International Solvay Institutes, Pleinlaan 2, B-1050 Brussels, Belgium\\
\email{hzhang@vub.ac.be}}

\abstract{In this note, we have addressed various issues on zero temperature holographic superfluids.
First, inspired by our numerical evidence for the equality between
 the superfluid density and particle density, we provide an elegant analytic proof for this equality by a boost trick.
 Second, using not only the frequency domain analysis but also the time domain analysis from numerical relativity, we identify the hydrodynamic normal modes and calculate out the sound speed, which is shown to increase with the chemical potential and saturate to the value predicted by the conformal field theory in the large chemical potential limit. Third, the generic non-thermalization is demonstrated by the fully non-linear time evolution from a non-equilibrium state for our zero temperature holographic superfluid. Furthermore, a conserved Noether charge is proposed in support of this behavior.}
\begin{document}

\section{Introduction}
Quantum phase transition occurs when the ground state of a many-body system is driven by quantum fluctuations to take an abrupt change. Ultracold atoms in optical lattices provide a clean system to test those quantum phase transitions predicted by various theories. For example, the quantum critical behavior of ultracold cesium atoms in an optical lattice across the vacuum to superfluid transition has recently been observed by tuning the chemical potential\cite{ZHTC}. Among others, AdS/CFT correspondence provides a natural framework to study such quantum criticality. In particular, resonant to the above experiment, the quantum phase transition from the vacuum to superfluid has been implemented by holography for the first time in \cite{NRT}, where by tuning the chemical potential the AdS soliton,  dual to the vacuum phase, will undertake a quantum phase transition to a hairy AdS soliton, which corresponds to a superfluid phase. Moreover, the compactified dimension in the AdS soliton background can be naturally identified as the reduced dimension in optical lattices by the very steep harmonic potential as both mechanisms make the effective dimension of the system in consideration reduced in the low energy regime.

The purpose of this note is to fill some gaps in the previous investigations of the above holographic model\footnote{Part of the results for the standard quantization have been presented in \cite{GNTZ} to demonstrate how to apply AdS/CFT with numerics, which is the emphasis of \cite{GNTZ}. For completeness, we include them into this note. } . Albeit well expected, the equality between the superfluid density and particle density remains an open question due to their different origins. Namely, the particle density can be read off directly from the static background while one is required to extract the superfluid density from the optical conductivity by the linear response theory on top of the static background. Inspired by our numerical evidence, we provide an elegant analytic proof for such an equality by a boost trick. In addition, apart from the frequency domain analysis of spectrum of normal modes, we introduce an alternative method to applied AdS/CFT from numerical relativity, namely the time domain analysis. To make this alternative work, we are required first to massage the equations of motion in terms of Hamilton formalism. Using either approach, we further extract sound speed from the hydrodynamic normal modes for our zero temperature holographic superfluid. The resultant sound speed increases with the chemical potential, and saturates to $\frac{1}{\sqrt{2}}$ or $\frac{1}{\sqrt{3}}$ in the large chemical potential limit, depending on whether the conformal dimension of condensate is $2$ or $1$. These two saturated values are consistent with those predicted by conformal field theory. On the other hand, we also exploit the Hamilton formalism to perform a fully non-linear temporal evolution, where the non-thermalization occurs generically. Inspired by this, we further provide a conserved Noether charge argument for this behavior.

The structure of this note is organized as follows. After a brief review of holographic model for zero temperature superfluids in Section \ref{casestudy}, we shall recall how to construct the corresponding phase diagram by the free energy analysis of background solutions in Section \ref{phasetransition}. Then in Section \ref{linearresponse}, we calculate the corresponding optical conductivity by the linear response theory and provide an analytic proof of the equality between the particle density and the superfluid density for our zero temperature holographic superfluid on top of the numerical evidence. In the subsequent section, using both the frequency and time domain analysis, we identify the hydrodynamic normal modes, read the sound speed out of the dispersion relation, and figure out the variation of sound speed with respect to the chemical potential. In Section \ref{nonthermalization}, not only do we perform a fully non-linear time evolution from the non-equilibrium state to demonstrate the non-thermalization, but also propose a conserved Noether charge to argue for this generic phenomenon. We end up with some discussions in the end.

\section{Holographic Model for Zero Temperature Superfluids}\label{casestudy}
In this section, we would like to present the holographic setup for zero temperature superfluids. The action for the simplest model of holographic superfluid is given by
\begin{equation} \label{action}
S=\frac{1}{16\pi G}\int d^{d+1}x\sqrt{-g}[R+\frac{d(d-1)}{l^2}+L_{matter}].
\end{equation}
Here $G$ is the Newton's gravitational constant, the AdS curvature radius is related to the negative cosmological constant as  $\Lambda=-\frac{d(d-1)}{2l^2}$, and  the Lagrangian for the matter fields reads
\begin{equation}
L_{matter}=\frac{l^2}{e^2}(-\frac{1}{4}F^{ab}F_{ab}-|D\Phi|^2-m^2|\Phi|^2)=\frac{l^2}{e^2}L,
\end{equation}
where $F=dA$, $D=\nabla-iA$, $e$ and $m$ are the charge and mass of complex scalar field.  To make our life much easier, we shall work with the probe limit, namely the back reaction of matter fields onto the metric is neglected, which can be achieved by taking the large $e$ limit. Thus we can put the matter fields on top of the background which is the solution to the vacuum Einstein equation with a negative cosmological constant. For our purpose, we can choose the AdS soliton as the bulk geometry\cite{NRT}, i.e.,
\begin{equation}
ds^2=\frac{l^2}{z^2}[-dt^2+d\mathbf{x}^2+\frac{dz^2}{f(z)}+f(z)d\chi^2].
\end{equation}
Here $f(z)=1-(\frac{z}{z_0})^d$ with $z=z_0$ the tip where our geometry caps off and $z=0$ the AdS boundary. To guarantee the smooth geometry at the tip, we are required to impose the periodicity $\frac{4\pi z_0}{d}$ onto the $\chi$ coordinate. The inverse of this periodicity set by $z_0$ is usually interpreted as the confining scale for the dual boundary theory. In what follows, we will take the units in which $l=1$, $16\pi G e^2=1$, and $z_0=1$. In addition, we shall focus solely on the action of matter fields because the leading $e^0$ contribution has been frozen by the above fixed background geometry.

The variation of action gives rise to the equations of motion as 
\begin{eqnarray}
%&&G_{ab}-\frac{d(d-1)}{2l^2}g_{ab}=\frac{l^2}{Q^2}[F_{ac}F_b{}^c+2D_a\Phi D_b\Phi-(\frac{1}{4}F_{cd}F^{cd}+|D\Phi|^2+m^2|\Phi|^2)g_{ab}],\\
&&\nabla_aF^{ab}=i(\overline{\Phi}D^b\Phi-\Phi\overline{D^b\Phi}),\\
&&D_aD^a\Phi-m^2\Phi=0,
\end{eqnarray}
whereby the asymptotical behavior for the bulk fields near the AdS boundary goes as follows
\begin{eqnarray}\label{conformaldimension}
%&&ds^2\rightarrow\frac{l^2}{z^2}[dz^2+(\gamma_{\mu\nu}+t_{\mu\nu}z^d)dx^\mu dx^\nu],\\
&&A_\mu\rightarrow a_\mu+b_\mu z^{d-2},\label{asympt1}\\
&&\Phi\rightarrow\phi_-z^{\Delta_-}+\phi_+z^{\Delta_+}\label{asympt2}
\end{eqnarray}
with the axial gauge $A_z=0$ and $\Delta_\pm=\frac{d}{2}\pm\sqrt{\frac{d^2}{4}+m^2}$. Below we shall concentrate ourselves onto the case of $d=3$ and $m^2=-2$. Consequently $\Delta_-=1$ and $\Delta_+=2$. Then by the holographic dictionary, we have

\begin{eqnarray}
\langle j^\mu\rangle&=&\frac{\delta S_{\mp}}{\delta a_\mu}=b^\mu,\nonumber\\
\langle O_\pm\rangle&=&\frac{\delta S_{\mp}}{\delta \phi_\mp}=\pm\overline{\phi_\pm}.
\end{eqnarray}
Here $j^\mu$ corresponds to the conserved particle current, the expectation value for the
scalar operator $O_\pm$ is interpreted as the condensate order parameter of our holographic superfluid, and $S_{\mp}$ is the holographic renormalized action by adding the counter terms to the original action to make it finite, given by
\begin{equation}
S_-=S-\int d^3x\sqrt{-h}|\Phi|^2, S_+=S+(\int d^3x\sqrt{-h}n_a\overline{D^a\Phi}\Phi+C.C.)+\int d^3x\sqrt{-h}|\Phi|^2,
 \end{equation}
depending on the choice of standard or alternative quantization, namely the choice of source $\phi_\mp$.
When this scalar operator carries a nonzero expectation value spontaneously in the situation where the source is turned off, the boundary system is driven into a superfluid phase. Now let us recall how to explicitly implement such a superfluid phase by our holographic model.

\section{Background Solutions, Free Energy, and Phase Transition}\label{phasetransition}
\begin{figure}
\begin{center}
\includegraphics[width=7.5cm]{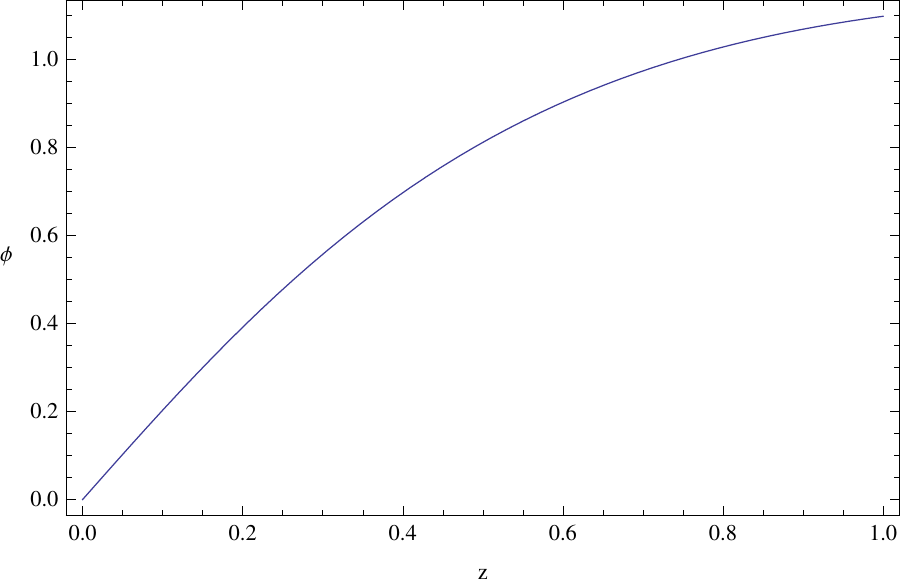}
\includegraphics[width=7.5cm]{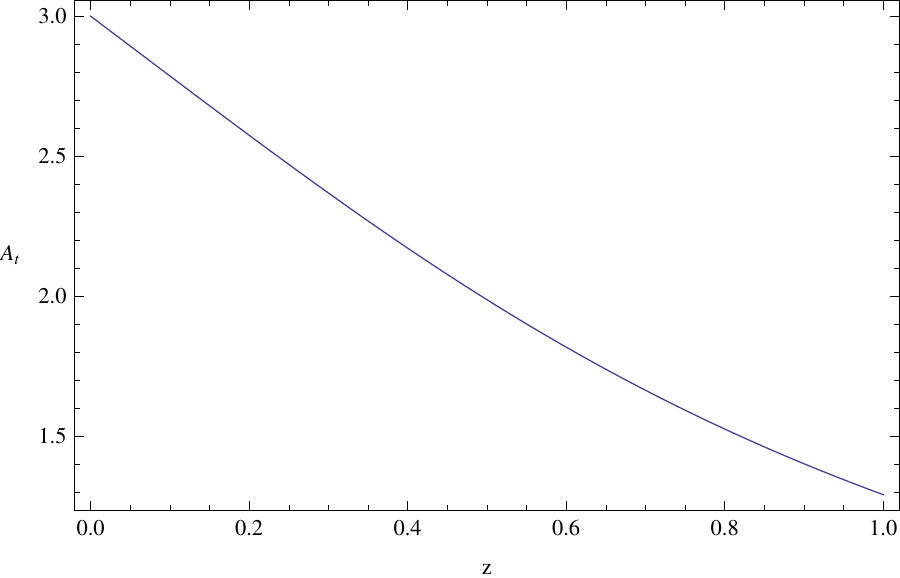}
\includegraphics[width=7.5cm]{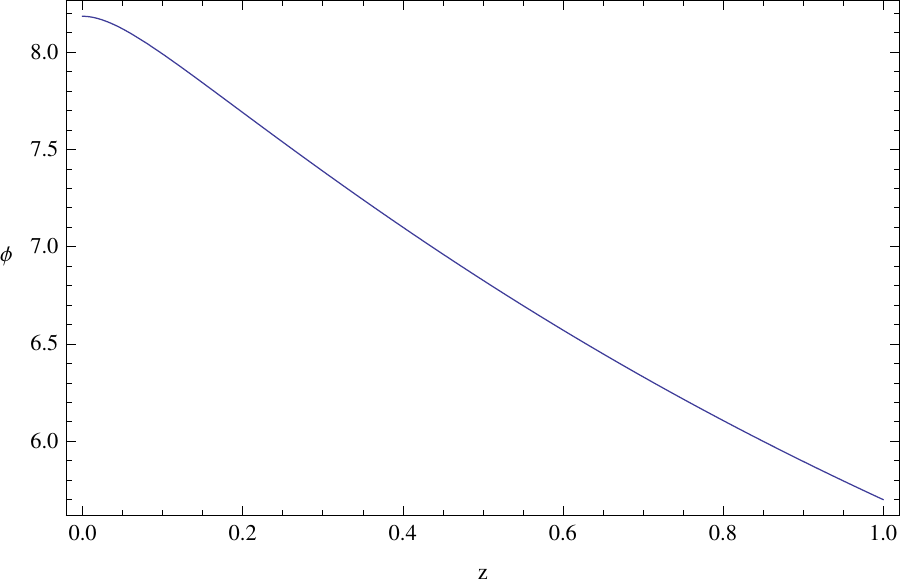}
\includegraphics[width=7.5cm]{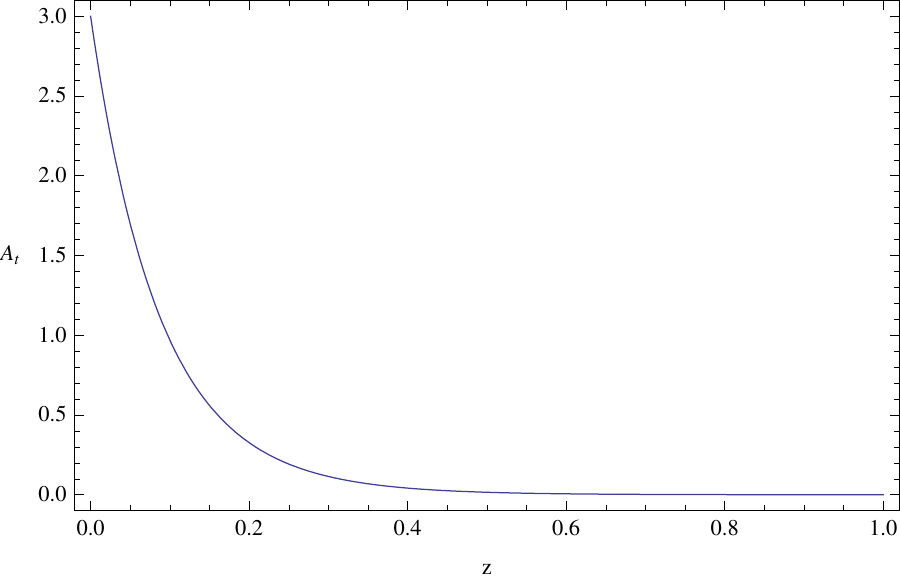}
\end{center}
\caption{The bulk profile for the scalar field and time component of gauge field at the chemical potential $\mu=3$, where the top is for the standard quantization, and the bottom is for the alternative quantization.}
\label{profile}
\end{figure}
\begin{figure}
\begin{center}
\includegraphics[width=7.0cm]{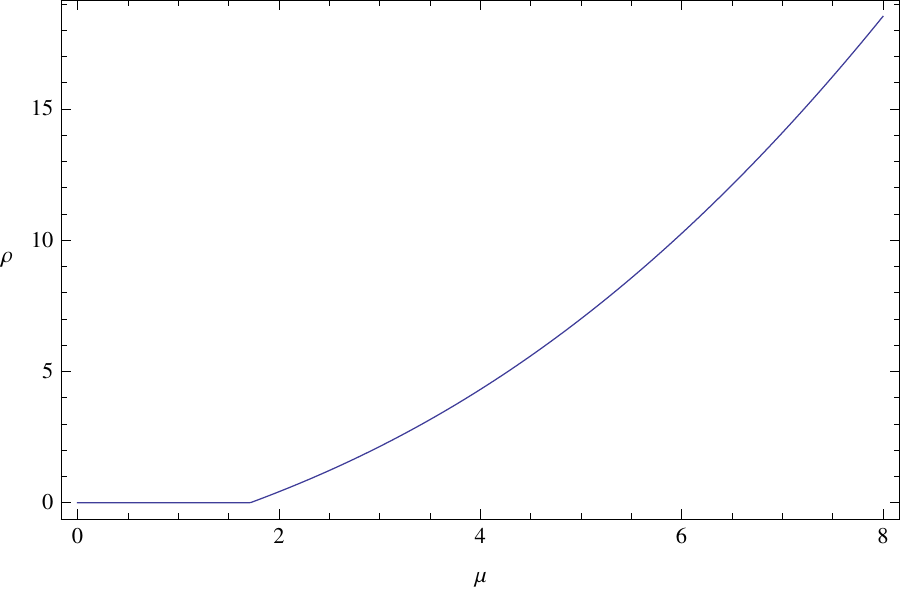}
\includegraphics[width=7.5cm]{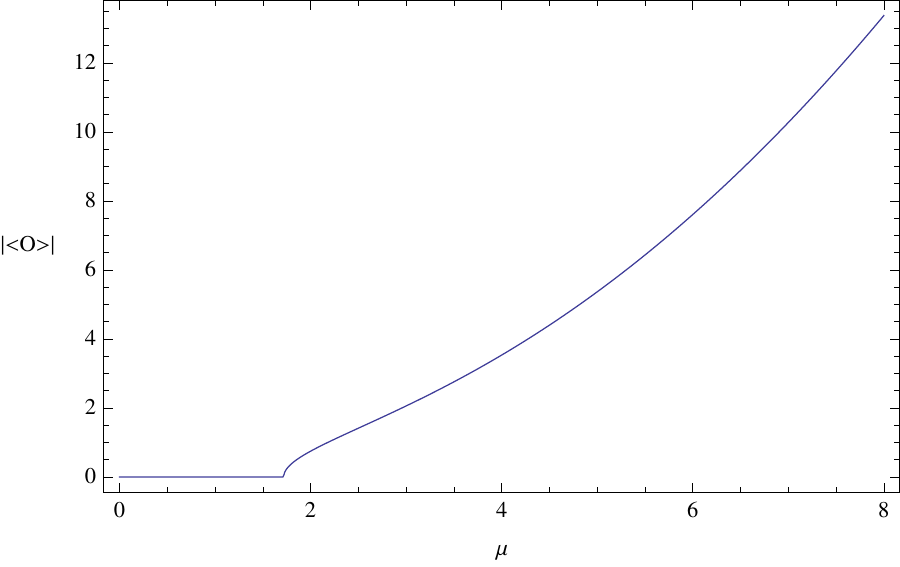}
\includegraphics[width=7.0cm]{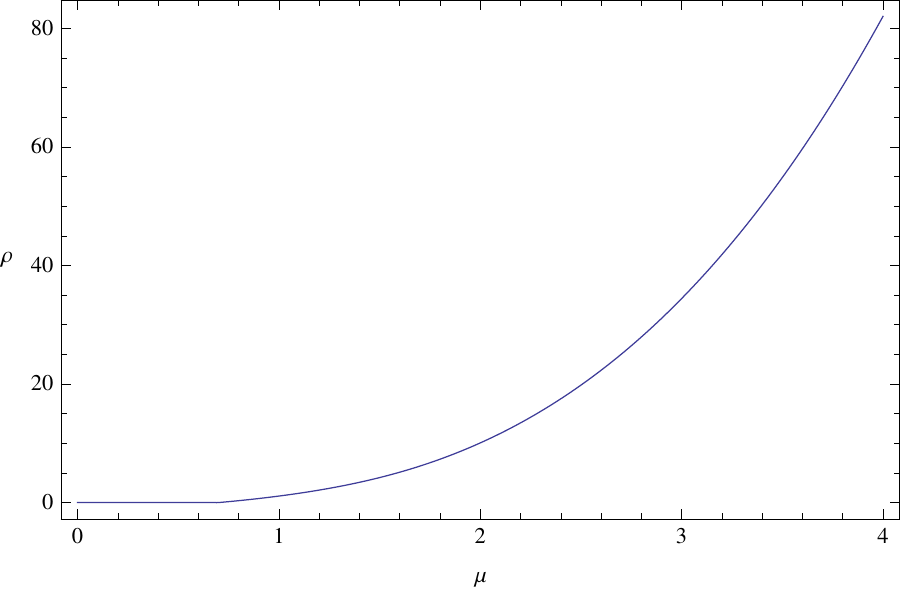}
\includegraphics[width=7.5cm]{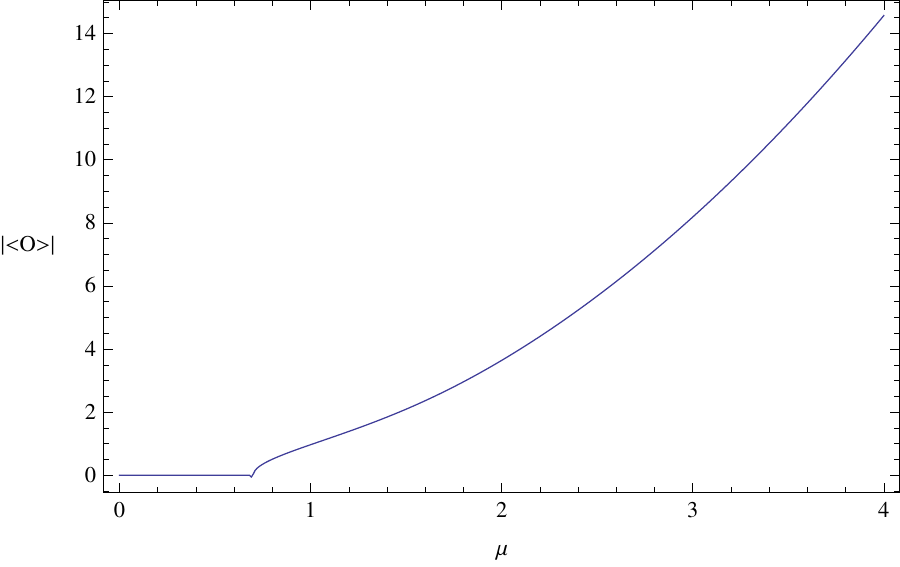}

\end{center}
\caption{The variation of particle density and condensate with respect to the chemical potential, where one can see the second order quantum phase transition take place at $\mu_c\approx1.715$ and $\mu_c\approx0.685$ for the standard quantization in the top and the alternative quantization in the bottom, respectively.}
\label{phase}
\end{figure}

For simplicity, we assume that the non-vanishing bulk matter fields $(\Phi=z\phi,A_t,A_x)$ do not depend on the coordinate $\chi$, then the equations of motion can be explicitly written as
\begin{eqnarray}
0&=&\partial_t^2\phi+(z+A_x^2-A_t^2+i\partial_xA_x-i\partial_tA_t)\phi+2iA_x\partial_x\phi-2iA_t\partial_t\phi-\partial_x^2\phi\nonumber\\
&&+3z^2\partial_z\phi+(z^3-1)\partial_z^2\phi,\\
0&=&\partial_t^2A_x-\partial_t\partial_xA_t-i(\phi\partial_x\bar{\phi}-\bar{\phi}\partial_x\phi)+2A_x\phi\bar{\phi}+3z^2\partial_zA_x+(z^3-1)\partial_z^2A_x,\\
0&=&(z^3-1)\partial_z^2A_t+3z^2\partial_zA_t-\partial_x^2A_t+\partial_t\partial_xA_x+2\bar{\phi}\phi A_t+i(\bar{\phi}\partial_t\phi-\phi\partial_t\bar{\phi}),\\
0&=&\partial_t\partial_zA_t+i(\phi\partial_z\bar{\phi}-\bar{\phi}\partial_z\phi)-\partial_z\partial_xA_x, \label{redundant}
\end{eqnarray}
where the third one is the constraint equation.
%and the last one reduces to the conserved equation for the boundary current when evaluated at the AdS boundary, i.e.,
%\begin{equation}
%\partial_t\rho=-\partial_xj^x.
%\end{equation}
To specialize into the homogeneous phase diagram for our holographic model\footnote{For the striped phase, please refer to \cite{EGP,KWG} for discussions.}, we further make the following ansatz for our non-vanishing bulk matter fields
\begin{equation}
\phi=\phi(z), A_t=A_t(z).
\end{equation}
Then the equations of motion for the static solution reduce to
\begin{eqnarray}
0&=&3z^2\partial_z\phi+(z^3-1)\partial_z^2\phi+(z-A_t^2)\phi,\\
0&=&2A_t\phi\bar{\phi}+3z^2\partial_zA_t+(z^3-1)\partial_z^2A_t,\\
0&=&\phi\partial_z\bar{\phi}-\bar{\phi}\partial_z\phi,
\end{eqnarray}
where the last equation implies that we can always choose a gauge to make $\phi$ real. Furthermore, it is not hard to see the above equations of motion have a trivial solution
\begin{equation}
\phi=0, A_t=\mu,
\end{equation}
which corresponds to the vacuum phase with zero particle density at the chemical potential $\mu$. On the other hand, as alluded to in the very beginning, when one cranks up the chemical potential there also exists a non-trivial solution dual to the superfluid phase, which can be obtained numerically by the pseudo-spectral method.
As a demonstration, we here plot
the nontrivial profile of $\phi$ and $A_t$ at $\mu=3$ for both quantizations in Figure \ref{profile}. The variation of particle density and condensate with respect to the chemical potential is plotted in Figure \ref{phase}, which indicates that  the phase transition from the vacuum to superfluid occurs at $\mu_c\approx1.715$ for the standard quantization and at $\mu_c\approx0.685$ for the alternative quantization. It is noteworthy that the particle density shows up at the same time as our superfluid condensate, thus it is tempting to suspect that this particle density $\rho$ is simply the superfluid density $\rho_s$. Actually this suspicion is also consistent with the fact that we are working with a zero temperature superfluid where all the normal fluid component should have been driven into the superfluid state. As we will show later on by the linear response theory, the superfluid density extracted from the optical conductivity is exactly equal to the particle density.

\begin{figure}
\begin{center}
\includegraphics[width=7.5cm]{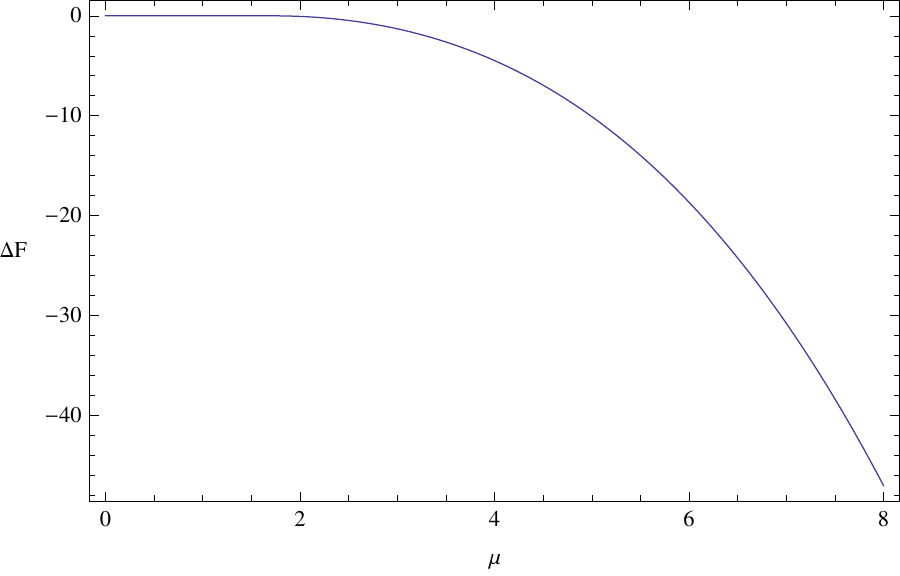}
\includegraphics[width=7.5cm]{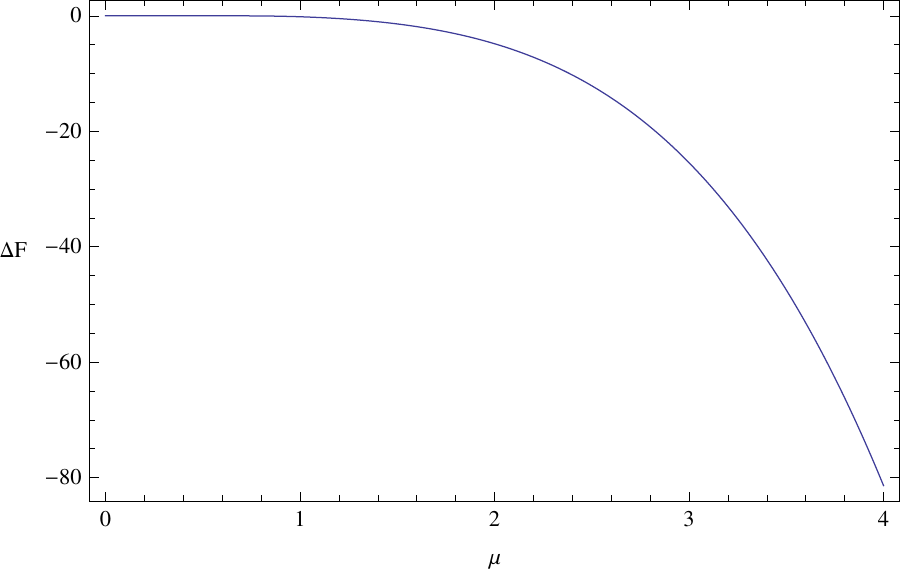}
\end{center}
\caption{The difference of free energy density for the superfluid phase from that for the vacuum phase, where the left is for the standard quantization and the right is for the alternative quantization.}
\label{energy}
\end{figure}

But we are left to ensure that Figure \ref{phase} represents the genuine phase diagram for our holographic model. As such, we are required to check whether the corresponding free energy density is the lowest in the grand canonical ensemble. By holography, the free energy density can be obtained from the renormalized on shell Lagrangian of matter fields as follows\footnote{Here we have used $iS_{Lorentzian}=-S_{Euclidean}=-\frac{FV}{T}$ with $V$ the volume of boundary system and $it=\tau$ with the Euclidean time $\tau$ identified as the inverse of temperature $T$.}
\begin{eqnarray}\label{free}
F&=&-\frac{1}{2}[\int dz \sqrt{-g}i(\overline{\Phi}D^b\Phi-\Phi\overline{D^b\Phi})A_b-\sqrt{-h}n_aA_bF^{ab}|_{z=0}]\nonumber\\
&=&-\frac{1}{2}\mu\rho+\int dz(A_t\phi)^2,
\end{eqnarray}
where we have made use of the equations of motion and the source free boundary condition for the scalar field at the AdS boundary. As expected in Figure
\ref{energy}, the superfluid phase is the thermodynamically favored one compared to the vacuum phase when the chemical potential is larger than the critical value.

\section{Linear Response Theory, Optical Conductivity, and Superfluid Density}\label{linearresponse}
Now let us set up the linear response theory on top of the above background solutions for the later calculation of the optical conductivity and normal modes  of our holographic model. To achieve this, we first decompose the field $\phi$ into its real and imaginary parts as
\begin{equation}
\phi=\phi_r+i\phi_i,
\end{equation}
and require that the perturbation bulk fields take the following form
\begin{equation}
\delta\phi_r=\delta\phi_r(z)e^{-i\omega t+iqx},\delta\phi_i=\delta\phi_i(z)e^{-i\omega t+iqx},\delta A_t=\delta A_t(z)e^{-i\omega t+iqx}, \delta A_x=\delta A_x(z)e^{-i\omega t+iqx},
\end{equation}
since the background solutions are static and homogeneous. With this,  the perturbation equations can be simplified as
\begin{eqnarray}
0&=&-\omega^2\delta\phi_r+(z-A_t^2)\delta\phi_r-2i\omega A_t\delta\phi_i+q^2\delta\phi_r+3z^2\partial_z\delta\phi_r+(z^3-1)\partial_z^2\delta\phi_r\nonumber\\
&&-2A_t\phi_r\delta A_t,\\
0&=&-\omega^2\delta\phi_i+(z-A_t^2)\delta\phi_i+2i\omega A_t\delta\phi_r+q^2\delta\phi_i+3z^2\partial_z\delta\phi_i+(z^3-1)\partial_z^2\delta\phi_i
\nonumber\\
&&+i\omega\phi_r\delta A_t+iq\phi_r\delta A_x,\\
0&=&-\omega^2\delta A_x-\omega q\delta A_t+3z^2\partial_z\delta A_x+(z^3-1)\partial_z^2\delta A_x+2\phi_r^2\delta A_x-2iq\phi_r\delta\phi_i,\label{opp}\\
0&=&(z^3-1)\partial_z^2\delta A_t+3z^2\partial_z\delta A_t+q^2\delta A_t+\omega q\delta A_x+2\phi_r^2\delta A_t+4A_t\phi_r\delta\phi_r\nonumber\\
&&+2i\omega\phi_r\delta\phi_i,\\
0&=&-i\omega\partial_z\delta A_t-iq\partial_z\delta A_x-2(\partial_z\phi_r\delta\phi_i-\phi_r\partial_z\delta\phi_i),\label{constraint}
\end{eqnarray}
where we have used $\phi_i=0$ for the background solution.

Note that the gauge transformation
\begin{equation}
A\rightarrow A+\nabla\theta, \phi\rightarrow\phi e^{i\theta}
\end{equation}
with
\begin{equation}
\theta=\frac{1}{i}\lambda e^{-i\omega t+iqx}
\end{equation}
generates a spurious solution to the above perturbation equations as follows
\begin{equation}
\delta A_t=-\lambda\omega, \delta A_x=\lambda q, \delta\phi=\lambda\phi.
\end{equation}
Such a redundancy can be removed by requiring $\delta A_t=0$ at the AdS boundary\footnote{The only exception is the $\omega=0$ case, which can always be separately nailed down if necessary.}. In addition, Dirichlet or Neumann boundary condition will be implemented on $\delta\phi$ at the AdS boundary depending on whether we are working with the standard or alternative quantization. On the other hand, taking into account the fact that the perturbation equation (\ref{constraint}) will be automatically satisfied in the whole bulk once the other perturbations are satisfied\footnote{This result comes from the following two facts.  One is related to Bianchi identity $0=\nabla_av^a=\frac{1}{\sqrt{-g}}\partial_\mu(\sqrt{-g}v^\mu)$ for Maxwell equation, whereby the $z$ component of Maxwell equation satisfies $\partial_z(\frac{v^z}{z^4})=0$ if the rest equations of motion hold. The other is special to our holographic model, in which one can show that the $z$ component of Maxwell equation turns out to be satisfied automatically at $z=1$ if the rest equations hold there.}, we can throw away (\ref{constraint}) from now on.  

\begin{figure}
\begin{center}
\includegraphics[width=7.5cm]{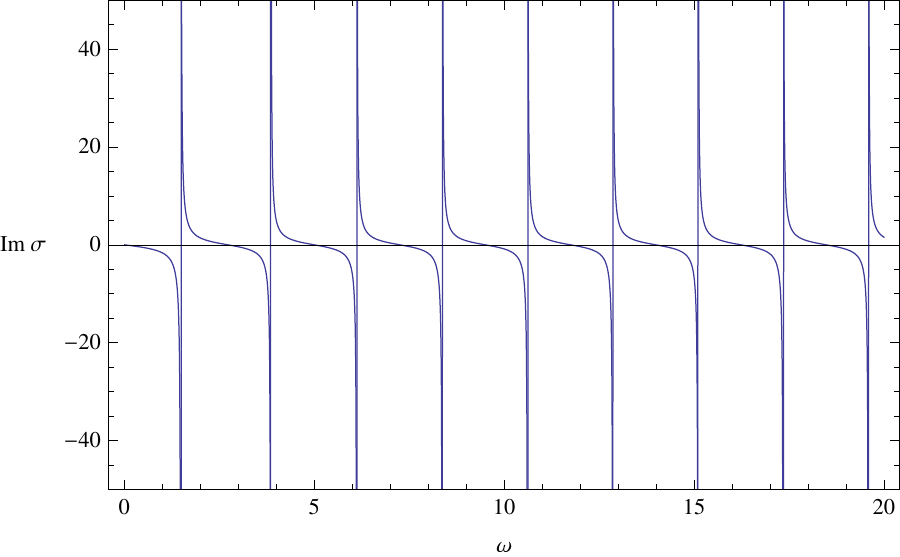}
\end{center}
\caption{The imaginary part of optical conductivity for the vacuum phase.}
\label{simple}
\end{figure}

\begin{figure}
\begin{center}

\includegraphics[width=7.5cm]{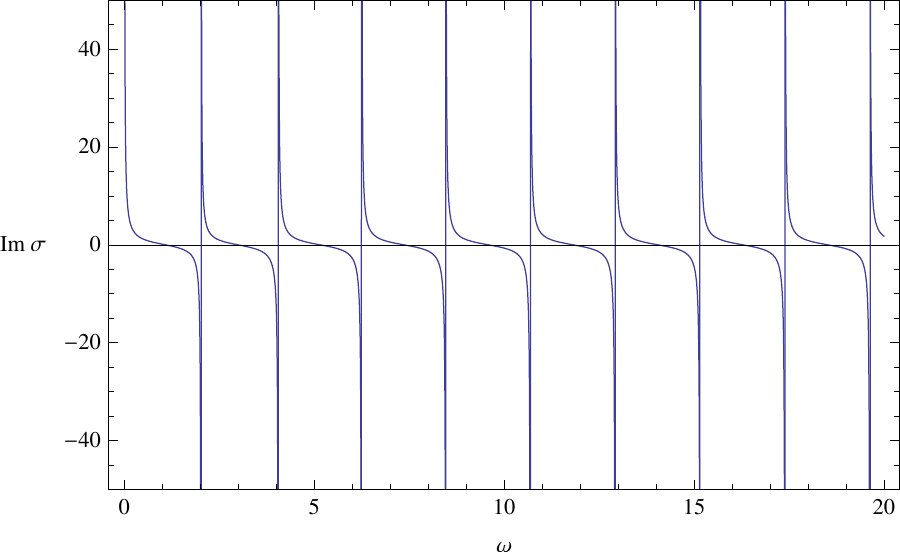}
\includegraphics[width=7.5cm]{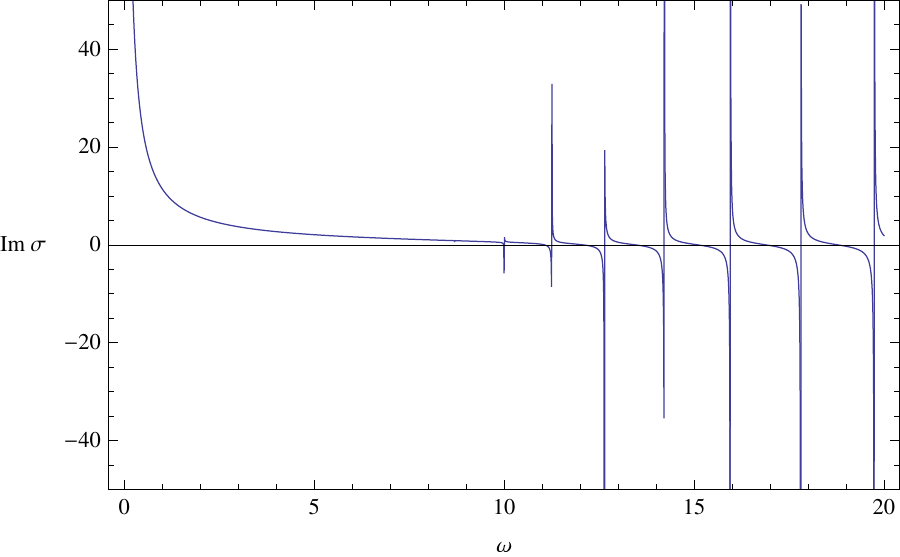}
\end{center}
\caption{The left panel is the imaginary part of optical conductivity for the superfluid phase at $\mu=3$ in the standard quantization, and the right panel is for the superfluid phase at the same chemical potential in the alternative quantization.}
\label{complex}
\end{figure}
In particular, to calculate the optical conductivity for our holographic model,  we can simply excite the $q=0$ mode. Accordingly $\delta A_x$ decouples from the other perturbation bulk fields, whereby we are required to deal exclusively with the perturbation equation (\ref{opp}) by the pseudo-spectral method.  With $\delta A_x=1$
at the AdS boundary,  the optical conductivity can be extracted by holography as
\begin{equation}\label{oc}
\sigma(\omega)=\frac{\partial_z\delta A_x|_{z=0}}{i\omega}
\end{equation}
for any positive frequency $\omega$\footnote{Taking into account $\sigma(-\bar{\omega})=\overline{\sigma(\omega)}$, we focus only on the positive frequency here.}. Because the real part vanishes due to the reality of the perturbation equation and boundary condition for $\delta A_x$, we simply plot the imaginary part of the optical conductivity in Figure \ref{simple} for the vacuum and in Figure \ref{complex} for our superfluid phase. As it should be the case, the DC conductivity vanishes for the vacuum phase, but diverges for the superfluid phase due to the $\frac{1}{\omega}$ behavior of the imaginary part of optical conductivity by the
Krames-Kronig relation
\begin{equation}
\mathbf{Im}[\sigma(\omega)]=\frac{1}{\pi}\mathcal{P}\int_{-\infty}^\infty d\omega'\frac{\mathbf{Re}[\sigma(\omega')]}{\omega-\omega'}.
\end{equation}
Furthermore, according to the hydrodynamical description of superfluid, the superfluid density $\rho_s$ can be obtained by fitting this zero pole as $\frac{\rho_s}{\mu\omega}$\cite{HKS,Yarom1,HY}.
As expected, our numerics shows that the resultant superfluid density is exactly the same as the particle density within our numerical accuracy. The other poles correspond to the gapped normal modes for $\delta A_x$, about which we are not caring since we are focusing on the low energy physics.

Let us come back to the equality between the particle density and superfluid density. Although it is 100 percent reasonable from the aforementioned physical perspective, this numerical result appears highly non-trivial in the sense that the superfluid density comes from the linear response theory while the particle density is a quantity associated with the equilibrium state. Such a numerical result, if not a coincidence,  begs an analytic understanding. Here we would like to develop an elegant proof for this equality by a boost trick. To this end, we are first required to realize $\rho_s=-\mu\partial_z\delta A_x|_{z=0}$ with $\omega=0$. Such an $\omega=0$ perturbation can actually be implemented by a boost
\begin{equation}
t=\frac{1}{\sqrt{1-v^2}}(t'-vx'), x=\frac{1}{\sqrt{1-v^2}}(x'-vt')
\end{equation}
acting on the superfluid phase. Note that the background metric is invariant under such a boost. As a result, we end up with a new non-trivial solution as follows
\begin{equation}
\phi'=\phi, A_t'=\frac{1}{\sqrt{1-v^2}}A_t, A_x'=-\frac{v}{\sqrt{1-v^2}}A_t.
\end{equation}
Expanding this solution up to the linear order in $v$, we have
\begin{equation}
\phi'=\phi, A_t'=A_t, A_x'=-vA_t.
\end{equation}
This means that the linear perturbation $\delta A_x$ is actually proportional to the background solution $A_t$, which gives rise to the expected equality $\rho_s=\rho$ immediately.

\section{Time Domain Analysis, Normal Modes, and Sound Speed}\label{timedomain}

\begin{figure}
\begin{center}
\includegraphics[width=7.5cm]{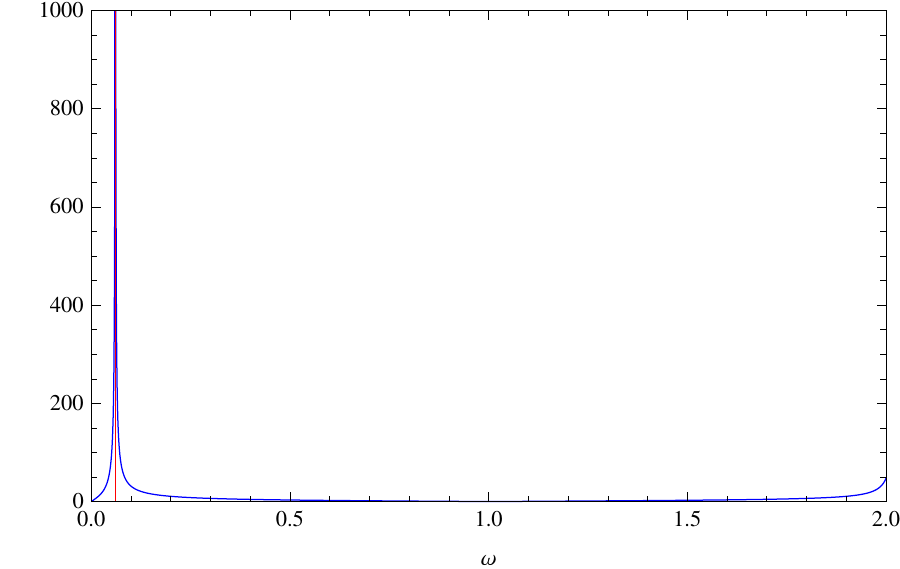}
\includegraphics[width=7.5cm]{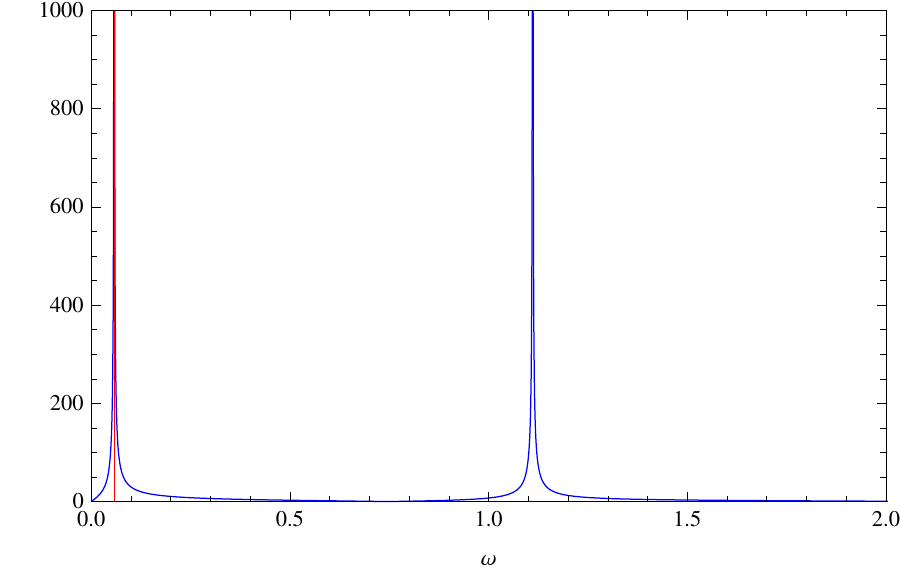}
\end{center}
\caption{ The density plot of $|\frac{det[\mathcal{L}(\omega)]'}{det[\mathcal{L}(\omega)]}|$ with $q=0.1$ for the superfluid phase at $\mu=3$, where the left is for the standard quantization and the right is for the alternative quantization. The normal modes can be identified by the peaks, where the red ones denote the hydrodynamic normal modes with $\omega_0\approx0.061$ for the standard quantization and $\omega_0\approx0.058$ for the alternative quantization.}
\label{densityplot}
\end{figure}

\begin{figure}
\begin{center}
\includegraphics[width=7.5cm]{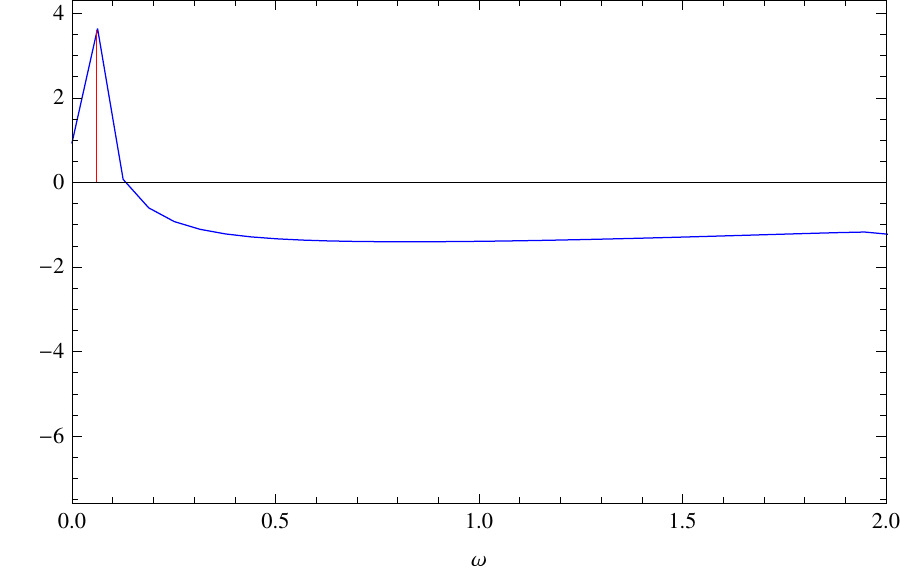}
\includegraphics[width=7.5cm]{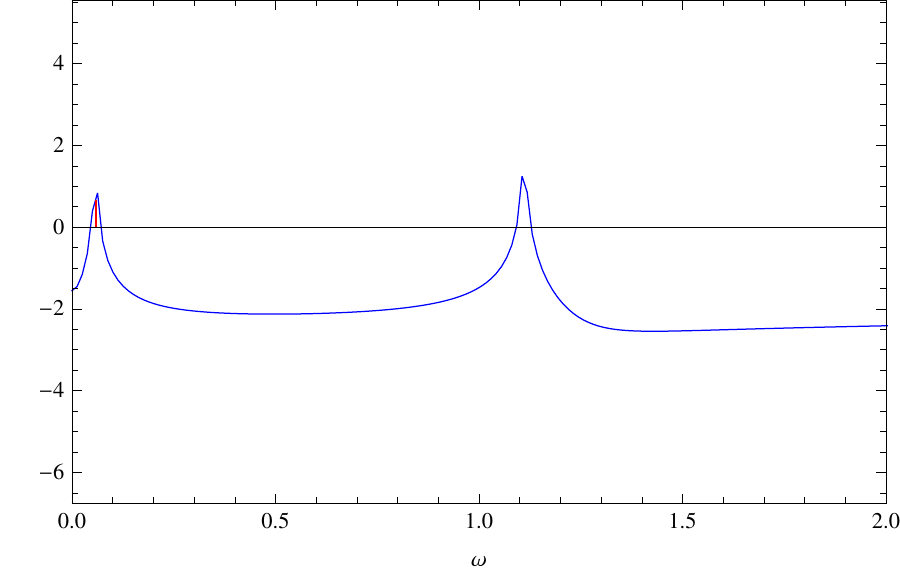}
\end{center}
\caption{ The spectral plot of $\ln{|\delta\hat{\phi}_i(\omega , 1)|}$ with $q=0.1$ for the superfluid phase at $\mu=3$, where the left is for the standard quantization and the right is for the alternative quantization.  We turn only on $\delta\phi_i$ at the initial time with $\delta\phi_i=z$ and $\delta\phi_i=z^2$ for the standard and alternative quantization respectively. The normal modes can be identified by the peaks, whose locations are the same as those by the frequency domain analysis within our numerical accuracy.}
\label{spectralplot}
\end{figure}
In this section, we shall use the linear response theory to calculate the speed of sound by focusing solely on the hydrodynamic spectrum of normal modes of the gapless Goldstone from the spontaneous symmetry breaking, which is obviously absent from the vacuum phase.  To achieve this, apart from the aforementioned boundary conditions specified for $\delta A_t$ and $\delta\phi$, we further impose Dirichlet boundary condition on $\delta A_x$ at the AdS boundary. Then the linear perturbation equations together with the boundary conditions can be cast into the form $\mathcal{L}(\omega)u= 0$ with $u$ the perturbation fields evaluated at the grid points by the pseudo-spectral method. The normal modes are obtained by the condition $det[\mathcal{L}(\omega)]= 0$, which can be further identified by the density plot  $|\frac{det[\mathcal{L}(\omega)]'}{det[\mathcal{L}(\omega)]}|$ with the prime the derivative with respect to $\omega$.  We demonstrate such a density plot in Figure \ref{densityplot}, where the hydrodynamic modes are simply the closest modes to the origin, marked in red. Besides such a frequency domain analysis of spectrum of normal modes, there is an alternative method called time domain analysis, which has been widely used in analyzing the black hole stability\cite{LTZZ}. We would like to take this opportunity to introduce this alternative method to the linear response theory in applied AdS/CFT. As such, we first cast the equations of motion into the following Hamilton formalism\footnote{When a black hole serves as the background, it is not necessary for one to cast the equations of motion into the Hamilton formalism by introducing some conjugate momenta. Instead, one can work in the Eddington coordinates, where the resulting equations of motion are first order in time by nature\cite{LTZZ,LTZ,DNTZ,DLTZ}.}
\begin{eqnarray}
\partial_t\phi&=&iA_t\phi+P,\\
\partial_tP&=&iA_tP-(z+A_x^2+i\partial_xA_x)\phi-2iA_x\partial_x\phi+\partial_x^2\phi-3z^2\partial_z\phi+(1-z^3)\partial_z^2\phi,\\
\partial_tA_x&=&\Pi_x+\partial_xA_t,\\
\partial_t\Pi_x&=&i(\phi\partial_x\bar{\phi}-\bar{\phi}\partial_x\phi)-2A_x\phi\bar{\phi}-3z^2\partial_zA_x+(1-z^3)\partial_z^2A_x,\\
0&=&(z^3-1)\partial_z^2A_t+3z^2\partial_zA_t+\partial_x\Pi_x-i(\bar{P}\phi-P\bar{\phi}),\\
\partial_t\partial_zA_t&=&-i(\phi\partial_z\bar{\phi}-\bar{\phi}\partial_z\phi)+\partial_z\partial_xA_x.
\end{eqnarray}
Then with the assumption that  the perturbation bulk fields take the form as $\delta(t,z)e^{iqx}$, the perturbation equations on top of the superfluid phase is given by
\begin{eqnarray}
\partial_t\delta\phi_r&=&-A_t\delta\phi_i+\delta P_r,\\
\partial_t\delta\phi_i&=&\phi_r\delta A_t+A_t\delta\phi_r+\delta P_i,\\
\partial_t\delta P_r&=&A_t\phi_r\delta A_t-A_t\delta P_i-(z+q^2)\delta\phi_r-3z^2\partial_z\delta\phi_r+(1-z^3)\partial_z^2\delta\phi_r,\\
\partial_t\delta P_i&=&-iq\phi_r\delta A_x+A_t\delta P_r-(z+q^2)\delta\phi_i-3z^2\partial_z\delta\phi_i+(1-z^3)\partial_z^2\delta\phi_i,\\
\partial_t\delta A_x&=&\delta\Pi_x+iq\delta A_t,\\
\partial_t\delta\Pi_x&=&2iq\phi_r\delta\phi_i-2\phi_r^2\delta A_x-3z^2\partial_z\delta A_x+(1-z^3)\partial_z^2\delta A_x,\\
0&=&(z^3-1)\partial_z^2\delta A_t+3z^2\partial_z\delta A_t+iq\delta\Pi_x-2\phi_r\delta P_i+2A_t\phi_r\delta\phi_r,\label{unique}\\
\partial_t\partial_z\delta A_t&=&2\partial_z\phi_r\delta\phi_i-2\phi_r\partial_z\delta\phi_i+iq\partial_z\delta A_x.
\end{eqnarray}
Supplemented with the previously prescribed boundary conditions for all the perturbation fields, we can obtain the temporal evolution of the perturbation fields for any given initial data by Runge-Kutta method, where $\delta A_t$ is solved by the constraint equation (\ref{unique}). The normal modes can then be identified by the peaks in the Fourier transformation of the evolving data. We demonstrate such a spectral plot in Figure \ref{spectralplot}. As expected, such a time domain analysis gives rise to the same result for the locations of normal modes as that by the frequency domain analysis.

\begin{figure}
\begin{center}
\includegraphics[width=7.5cm]{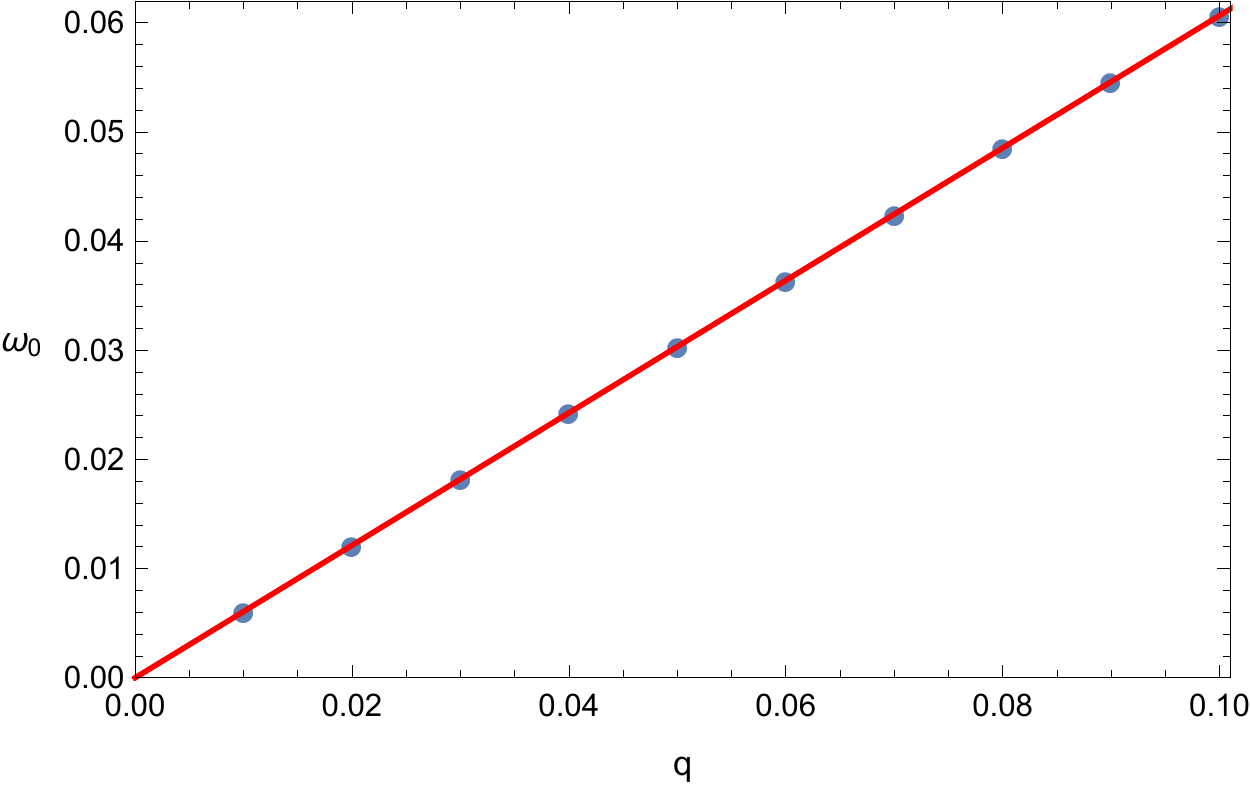}
\includegraphics[width=7.5cm]{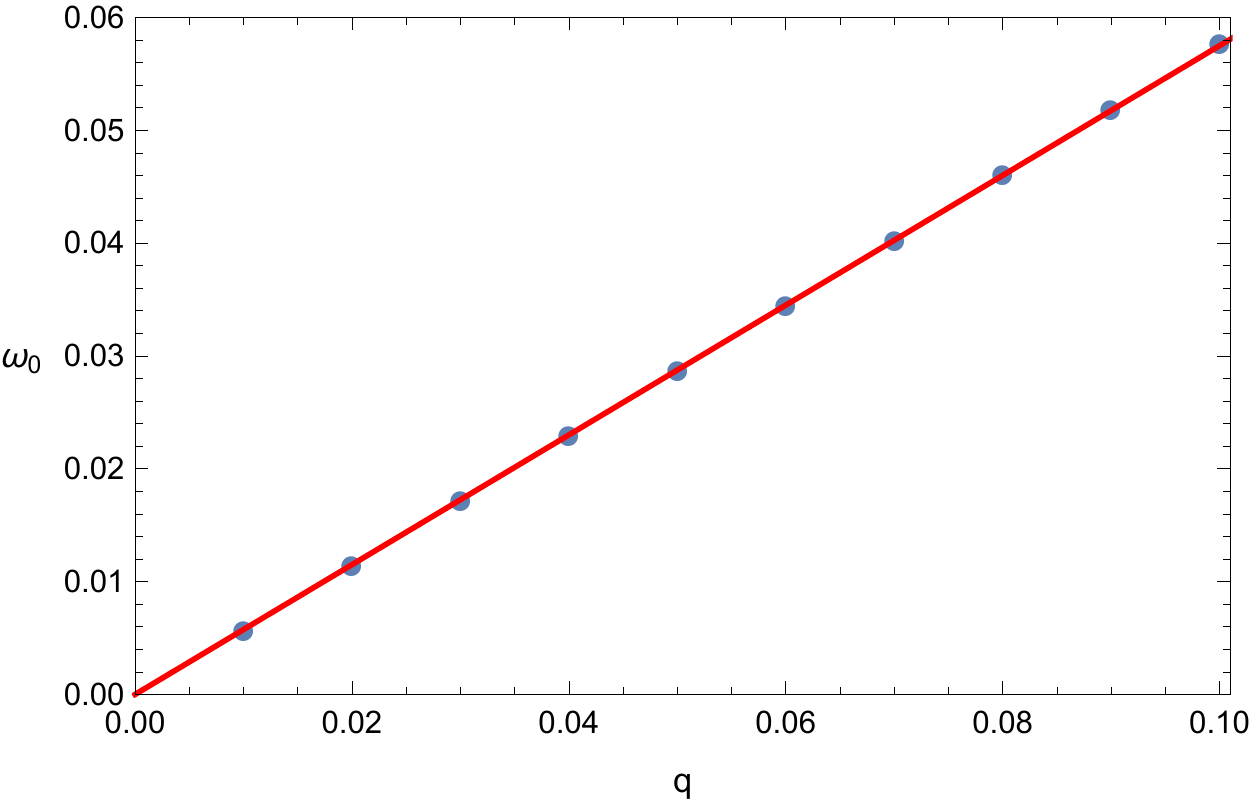}
\end{center}
\caption{ The dispersion relation for the gapless Goldstone mode in the superfluid phase at $\mu=3$. By fitting the long wave modes with $\omega_0=v_sq$, the sound speed is obtained as $v_s=0.606$ for the standard quantization on the left and $v_s=0.577$ for the alternative quantization on the right.}
\label{dispersion}
\end{figure}

\begin{figure}
\begin{center}
\includegraphics[width=7.5cm]{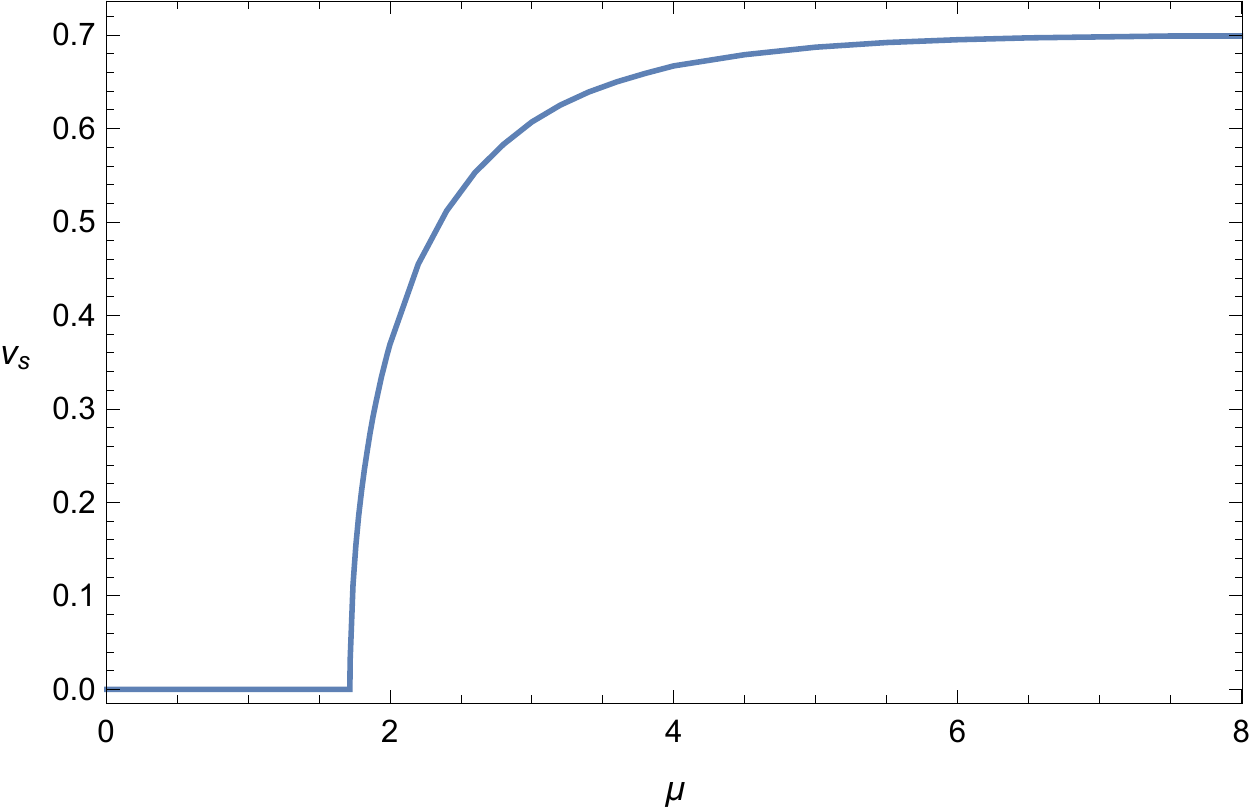}
\includegraphics[width=7.5cm]{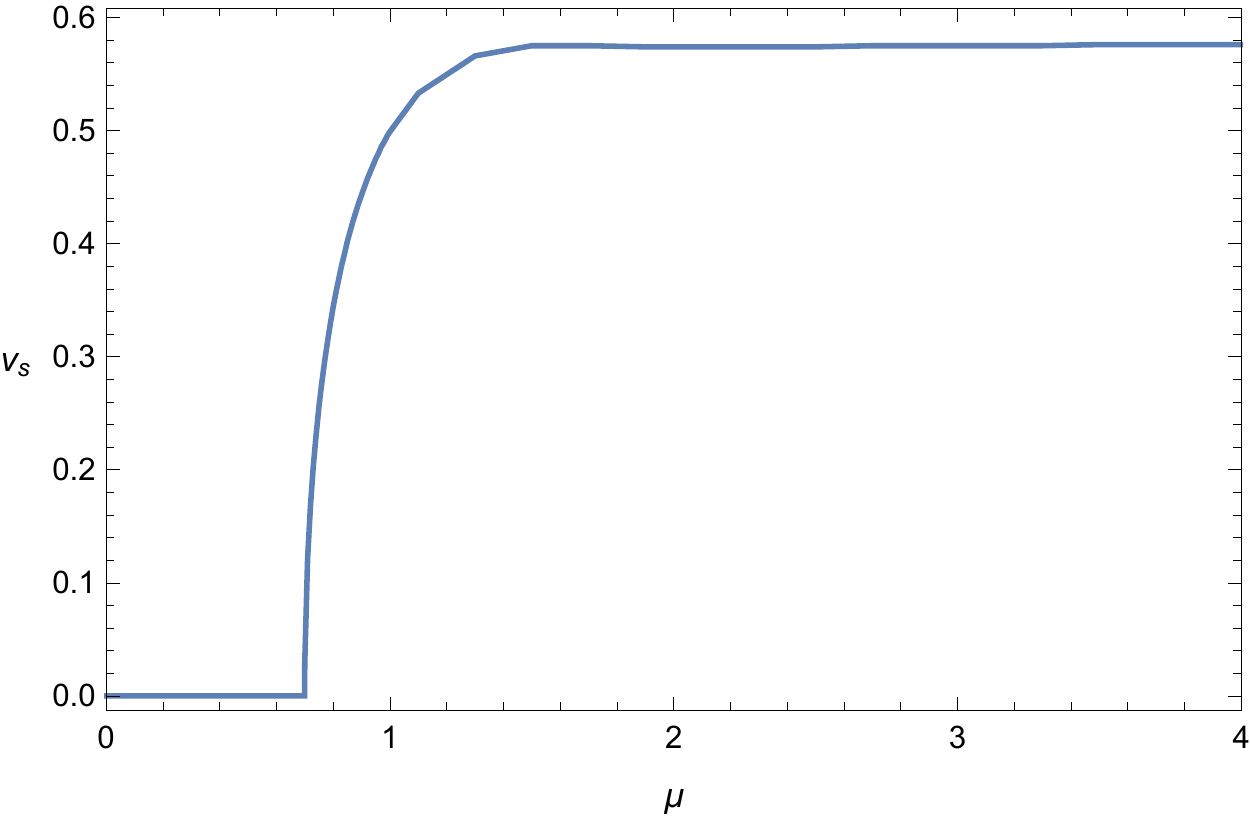}
\end{center}
\caption{ The variation of sound speed with respect to the chemical potential. When the chemical potential is much larger than the confining scale, the conformality is restored and the sound speed approaches the predicted value by conformal field theory: $\frac{1}{\sqrt{2}}$ for the standard quantization on the left and $\frac{1}{\sqrt{3}}$ for the alternative quantization on the right.}
\label{ss}
\end{figure}

Then the dispersion relation for the gapless Goldstone can be obtained and plotted in Figure \ref{dispersion}, whereby the sound speed $v_s$ can be further extracted by the fitting formula $\omega_0=v_s q$. As shown in Figure \ref{ss}, the sound speed increases with the chemical potential. Furthermore,  when the chemical potential is much larger than the confining scale, the sound speed saturates to the predicted value from conformal field theory, namely $\frac{1}{\sqrt{2}}$ for the standard quantization and $\frac{1}{\sqrt{3}}$ for the alternative quantization\cite{HKS,Yarom1,HY}.  This is reasonable since it is believed that the conformality is restored in this large chemical limit.

\section{Non-linear Evolution, Non-thermalization, and Conserved Charge Argument}\label{nonthermalization}
Generically, there are two ways to address the thermalization. One is to quench an equilibrium state and see what happens. The other is to start from a non-equilibrium state and see how it evolves. These two ways are essentially the same because one can always prepare such a non-equilibrium state by quench. But to our knowledge, the previous literature only takes the first way to investigate the holographic thermalization associated with the confining geometry. 
When the back reaction of matter fields onto the metric is taken into account, there are two scenarios, namely non-thermalization and thermalization, depending on whether the quench is weak or strong compared to the confining scale\cite{Brussels1,Brussels2,CET}. While in the probe limit, one ends up only with the non-thermalization, which has been observed for the zero temperature holographic superfluid in \cite{GGZZ}. In this section, we would like to take the second way to explore such a non-thermalization for our zero temperature holographic superfluid. 

\begin{figure}
\begin{center}
\includegraphics[width=7.5cm]{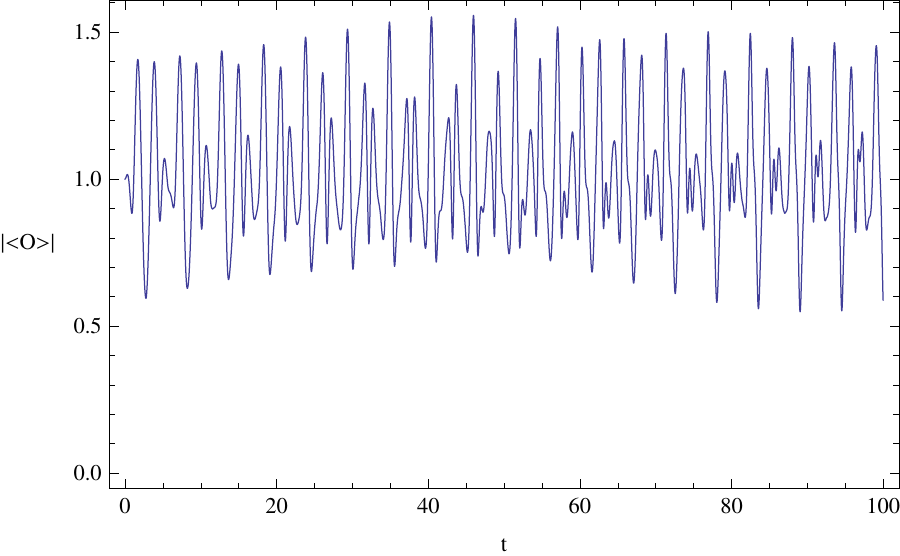}
\includegraphics[width=7.5cm]{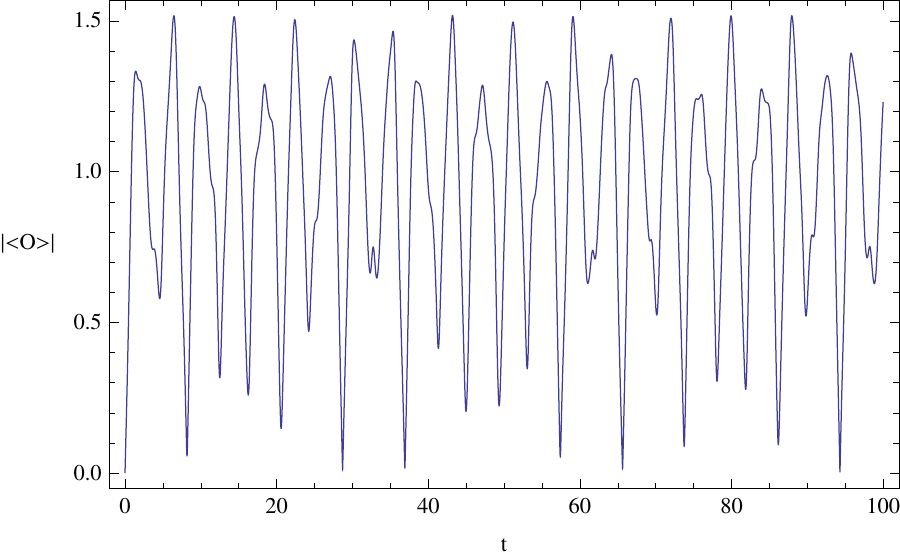}
\end{center}
\caption{ The temporal evolution of order parameter for the superfluid phase at $\mu=3$, where the left is for the standard quantization and the right is for the alternative quantization. }
\label{evolution}
\end{figure}
Different from the pseodu-spectral method used in \cite{GGZZ} for the temporal evolution, we shall exploit the previous Hamilton formalism to perform a fully non-linear time evolution by the Runge-Kutta method, which is supposed to be more efficient. We demonstrate such a temporal evolution of order parameter for both quantizations in Figure \ref{evolution}, where we prepare $\phi=z, P=iz, A_x=\Pi_x=0$  as the initial data for the standard quantization and $\phi=z^2, P=i, A_x=\Pi_x=0$ as the initial data for the alternative quantization. As illustrated in Figure \ref{evolution}, the order parameter keeps oscillating and does not tend to thermalize to an equilibrium state. Such a lack of thermalization is a generic phenomenon for the zero temperature holographic superfluid\cite{GGZZ}. So it is better to have an analytic understanding of this behavior. Inspired by the usual integrability argument for the non-thermalization, we propose a conserved bulk Noether charge associated with the timelike Killing vector field $\xi=\frac{\partial}{\partial t}$ in support of such a non-equilibration we run into right now. 

For the standard quantization, the conserved Noether current is given by\cite{Wald,TWZ0,TWZ1}
\begin{equation}
J^a=-F^{ab}\mathcal{L}_\xi A_b-(\overline{D^a\Phi}\mathcal{L}_\xi\Phi+C.C.)-\xi^aL.
\end{equation}
The flux across the AdS boundary vanishes due to the prescribed boundary condition for our temporal evolution. So the Noether charge defined by the flux across any surface of equal time is conserved. However, one can show that the flux across the AdS boundary by this conserved current diverges for the alternative quantization. But nevertheless, by adding to the original Lagrangian with a total derivative $\Delta L=\frac{1}{2}\nabla_a(\overline{D^a\Phi}\Phi+C.C.)$, we can redefine the conserved Noether current as follows
\begin{equation}
J'^a=J^a+\frac{1}{2}\mathcal{L}_\xi(\overline{D^a\Phi}\Phi+C.C.)-\xi^a\Delta L,
\end{equation}
which gives rise to a vanishing flux across the AdS boundary. So the corresponding Noether charge is conserved again.

Generically, the above Noether charge for a non-equilibrium state is not equal to that for the equilibrium state with the same boundary conditions at AdS boundary. So it is impossible for such a non-equilibrium state to equilibrate towards the would-be equilibrium state.

\section{Conclusion and Discussion}
After constructing the phase diagram for our zero temperature holographic superfluid by the free energy analysis of the background solutions, we make use of the linear response theory to work out the superfluid density and sound speed. In particular, inspired by the numerical evidence for the equality between the superfluid density and particle density, we provide an analytic proof for it by the boost trick. On the other hand, the resulting sound speed is found to increase with the chemical potential and saturate to the value predicted by conformal field theory for both quantizations. In passing, we introduce the time domain analysis to applied AdS/CFT, which gives rise to the same result on the spectrum of normal modes as the frequency domain analysis. Finally we perform a fully non-linear dynamical evolution for our zero temperature holographic superfluid to demonstrate the generic non-thermalization, which is further supported by the conserved Noether charge associated with the timelike Killing field.

We would like to conclude this note with two interesting issues worthy of further investigation. First, as suggested by the numerics in \cite{EHKMS}, the superfluid density is also equal to the particle density at zero temperature for the holographic superfluid in anisotropic holographic insulators, where the back reaction of matter fields onto the metric is taken into account. It is worthwhile to see whether our boost trick can also be applied to prove this equality for these situations. Second, one intriguing observation is that our Noether charge is exactly the same as the free energy for an equilibrium state, which may provide us with a new perspective into the correlated instability, namely the equivalence between the dynamical instability and thermodynamical instability. We hope to address this issue somewhere else\cite{TWZ2}.

\begin{acknowledgments}
H.Z. would like to thank the organizers of the Eleventh International Modave Summer School on Mathematical Physics held in Modave, Belgium, September 2015, where part of the results have been presented for the first time. He is indebted to Nabil Iqbal for his stimulating discussion on the equality between the particle density and superfluid density at the summer school.  H.Z. is also grateful to Rene Meyer for his valuable discussion on this equality and Amos Yarom for his helpful correspondence on the superfluid sound speed. In addition, he would like to acknowledge Ran Li for his stimulating discussion on the temporal evolution in the alternative quantization. M.G. is partially supported by NSFC with Grant Nos.11235003, 11375026 and NCET-12-0054. S.L. is supported by the NSFC with Grant No.11235003. C.N. is supported by Basic Science Research Program through the National Research Foundation of Korea(NRF) funded by the Ministry of Science, ICT \& Future Planning(NRF- 2014R1A1A1003220) and the 2015 GIST Grant for the FARE Project (Further Advancement of Research and Education at GIST College). Y.T.  is partially supported by NSFC with Grant No.11475179. H.Z. is supported in part by the Belgian Federal
Science Policy Office through the Interuniversity Attraction Pole
P7/37, by FWO-Vlaanderen through the project
G020714N, and by the Vrije Universiteit Brussel through the
Strategic Research Program ``High-Energy Physics''. He is also an individual FWO Fellow supported by 12G3515N.

\end{acknowledgments}


\begin{thebibliography}{20}
\bibitem{ZHTC} X. Zhang, C. L.  Hung, S. K. Tung, and C. Chin, Science 335, 1070(2012).
\bibitem{NRT}T. Nishioka, S. Ryu, and T. Takayanagi, JHEP 1003, 131(2010).
\bibitem{GNTZ}M. Guo, C. Niu, Y. Tian, and H. Zhang, arXiv:1601.00257.
\bibitem{EGP}J. Erdmenger, X. H. Ge, and D. W. Pang, JHEP 11, 027(2013).
\bibitem{KWG}X. M. Kuang, B. Wang, and X. H. Ge, Mod. Phys. Lett. A 29, 1450070(2014).
\bibitem{HKS}C. P. Herzog, P. K. Kovtun, and D. T. Son, Phys. Rev. D 79, 066002(2009).
\bibitem{Yarom1}A. Yarom,  JHEP 0907, 070(2009).
\bibitem{HY}C. P. Herzog and A. Yarom, Phys. Rev. D 80, 106002(2009).
%\bibitem{Lattice1}Y. Ling, C. Niu, J. P. Wu, Z. Xian, and H. Zhang, JHEP 07, 045(2013).
%\bibitem{Lattice2}Y. Ling, C. Niu, J. P. Wu, Z. Xian, and H. Zhang, Phys. Rev. Lett. 113, 091602(2014).
\bibitem{LTZZ}R. Li, Y. Tian, H. Zhang, and J. Zhao, Phys. Lett. B 750, 520(2015).
\bibitem{LTZ}W. J. Li, Y. Tian, and H. Zhang, JHEP 07, 030(2013).
\bibitem{DNTZ} Y. Du, C. Niu, Y. Tian, and H. Zhang, JHEP 12, 018(2015).
\bibitem{DLTZ}Y. Du, S. Q. Lan, Y. Tian, and H. Zhang, JHEP 01, 016(2016).
\bibitem{Brussels1}B. Craps, E. Kiritsis, C. Rosen, A. Taliotis, J. Vanhoof, and H. Zhang, JHEP 02, 120(2014).
\bibitem{Brussels2}B. Craps, E. J. Lindgren, A. Taliotis, J. Vanhoof, and H. Zhang, Phys. Rev. D 90, 086004(2014).
\bibitem{CET}B. Craps, E. J. Lindgren, and A. Taliotis, JHEP 12, 116(2015).
\bibitem{GGZZ}X. Gao, A. M. García-García, H. B. Zeng, and H. Q. Zhang, JHEP 06,  019(2014).
\bibitem{Wald}V. Iyer and R. Wald, Phys. Rev. D 50, 846(1994).
\bibitem{TWZ0}TY. Tian, X. N. Wu, and H. Zhang, Class.Quant.Grav. 30, 125010(2013).
\bibitem{TWZ1}Y. Tian, X. N. Wu, and H. Zhang, JHEP 1410, 170(2014).
\bibitem{EHKMS} J. Erdmenger, B. Herwerth, S. Klug, R. Meyer, and K. Schalm, JHEP 05, 094(2015).
\bibitem{TWZ2}Y. Tian, X. N. Wu, and H. Zhang, in preparation.
\end{thebibliography}
\end{document}